\documentclass[12pt]{article}

\newcommand{\bibl}[5]
	{#1, {\it #2} {{\bf #3},} #5 (#4)}
\newcommand{\cebe}{\begin{center}}
\newcommand{\ceen}{\end{center}}
\newcommand{\debe}{\begin{description} \vspace{-2ex}}
\newcommand{\deen}{\end{description}}
\newcommand{\eabe}{\begin{eqnarray}}
\newcommand{\eaen}{\end{eqnarray}}

\newcommand{\eqbe}{\begin{equation}}
\newcommand{\eqen}{\end{equation}}

\newcommand{\itbe}{\begin{itemize}}
\newcommand{\iten}{\end{itemize}}

\newcommand{\tabe}{\begin{tabbing}}
\newcommand{\taen}{\end{tabbing}}

\usepackage{psfig}

\begin{document}

\begin{titlepage}
\begin{flushright}
  LU TP 97-28 \\
  October 1997
\end{flushright}
\vspace{25mm}
\begin{center}
  \Large
  {\bf Transverse and Longitudinal \\ Bose-Einstein Correlations} \\
  \normalsize
  \vspace{12mm}
  Bo Andersson and
  Markus Ringn\'{e}r\footnote{E-mail: bo@thep.lu.se, markus@thep.lu.se}
  \vspace{1ex} \\
  Department of Theoretical Physics, Lund University, \\
  S\"olvegatan 14A, S-223 62 Lund, Sweden \\
\end{center}
\vspace{20mm}

\noindent {\bf Abstract:} \\ We show how a difference in the
correlation length longitudinally and transversely, with respect to
the jet axis in $e^+e^-$ annihilation, arises naturally in a model for
Bose-Einstein correlations based on the Lund string model. In genuine
three-particle correlations the difference is even more apparent and
they provide therefore a good probe for the longitudinal stretching of
the string field. The correlation length between pion pairs is found
to be rather independent of the pion multiplicity and the kaon content of
the final state.

\vspace{3cm}



\end{titlepage}

\section{Introduction} \vspace{-2ex}
The Hanbury-Brown-Twiss (HBT) effect (popularly known as the
Bose-Einstein effect) corresponds to an enhancement in the two
identical boson correlation function when the two particles have
similar energy-momenta. A well-known formula
\cite{r:bowler} to relate the two-particle correlation function (in four momenta
$p_j,j=1,2$ with $q=p_1-p_2$) to the space-time density, $\rho$, of
(chaotic) emission sources is
\begin{eqnarray}
\frac{\sigma d^2\sigma_{12}}{d\sigma_1 d\sigma_2}= 1 + |{\cal R}(q)|^2
\end{eqnarray}
where ${\cal R}$ is the normalised Fourier transform of the source density
\begin{eqnarray}
{\cal R}(q) = \frac{\int \rho(x) \exp(iqx)}{\int\rho(x) dx}
\end{eqnarray}
The commonly used event generators HERWIG and JETSET are based upon
classical stochastical processes and do not include HBT-effects
(although Sj\"ostrand, in JETSET, has introduced an
ingenious method to simulate any given distribution by means of a kind
of mean-field potential attraction between the bosons in the final
state).

In this letter we will further investigate some features of the
methods developed in \cite{r:bamr} (an extension of \cite{r:bawh} to
multi-boson final states). We will show that the model predicts, due to
the properties of string fragmentation, a difference between the
correlation length along the string and transverse to it. In
practice this means that if we introduce the longitudinal and
transverse components of the vector $q$ (defined with respect to the
thrust direction) then we obtain a noticeable difference in the
correlation distributions. This becomes even more noticeable when we go
to the three-particle HBT effect (which was predicted in \cite{r:bamr})
because in this case even more of the longitudinal stretching of the
string field becomes obvious. Finally we will investigate the
influence of the kaon and baryon content of the states on the HBT
effects between the pions.

\section{Longitudinal and transverse correlation lengths} 
\begin{figure}[t]
  \hbox{\vbox{
    \begin{center}
    \mbox{\psfig{figure=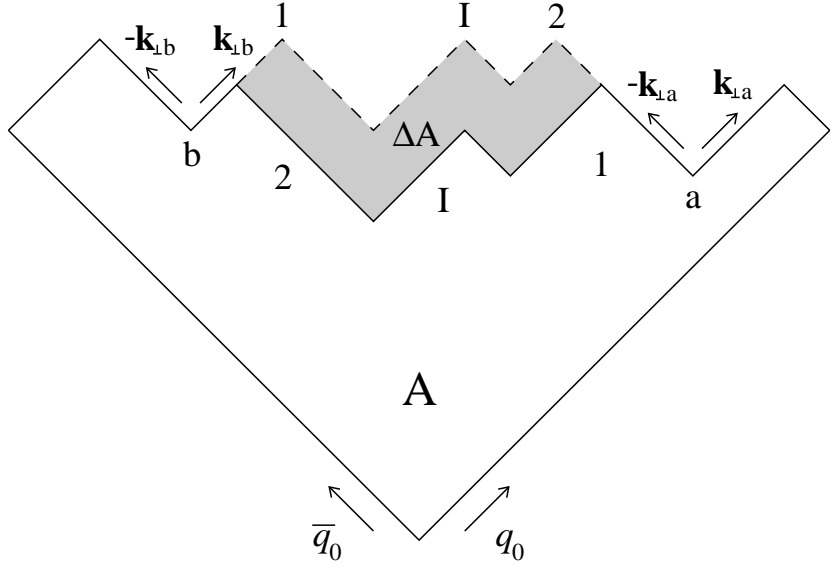,width=11cm}}
    \end{center}
  }}
\caption{\em The decay of a Lund Model string spanning the space--time 
area $A$. The particles $1$ and $2$ are identical bosons and the
particle(s) produced in between them is denoted by $I$. The two possible
ways, $(\ldots,1,I,2,\ldots)$ and $(\ldots,2,I,1,\ldots)$, to produce
the state are shown and the area difference between the two cases, $\Delta
A$, is shaded.  The two neighbouring vertices of the state with the two
identical bosons are denoted by $a$ and $b$, and the transverse momenta
of the quarks produced in the neighbouring vertices are $\pm {\bf
k}_{\perp a}$ and $\pm {\bf k}_{\perp b}$, respectively.}
\label{f:lundarea}
\end{figure}

\vspace{-2ex}
The starting point of our Bose-Einstein model \cite{r:bawh,r:bamr} is an
interpretation of the (non-normalised) Lund string area fragmentation
probability for an n-particle state (cf Fig \ref{f:lundarea})
\begin{eqnarray}
dP (p_1,p_2, \ldots,p_n)= \prod_1^n Ndp_j \delta(p_j^2-m_j^2)
\delta(\sum p_j-P_{tot}) \exp(-bA)
\label{e:lundprob}
\end{eqnarray}
in accordance with a quantum mechanical transition probability
containing the final state phase space multiplied by the square of a
matrix element ${\cal M}$. In \cite{r:bawh} and in more detail
in~\cite{r:bamr} a possible matrix element is suggested in accordance
with (Schwinger) tunneling and the (Wilson) loop operators necessary
to ensure gauge invariance. The matrix element is
\begin{eqnarray}
{\cal M}=\exp(i\kappa- b/2)A
\label{e:lundM}
\end{eqnarray}
where the area $A$ is interpreted in coordinate space, $\kappa$ is the
string constant (phenomenologically $\kappa \simeq 1$ $GeV/fm$) and
$b\simeq 0.3$ $GeV/fm$ is the decay constant. Note that the parameter $b$
is much smaller than $\kappa$.  From now on we will, as is usual in
the Lund model, go over to the energy momentum space. Then the area
$A\rightarrow 2 \kappa^2 A$, while $b\rightarrow b/2\kappa^2$, as
explained in~\cite{r:bamr}.

The transverse momentum properties are in the Lund model taken into account by
means of a Gaussian tunneling process. In this way the produced $(q
\bar{q})$-pair in each vertex will obtain $\pm {\bf k}_{\perp}$ and the hadron
stemming from the combination of a $\bar{q}$ from one vertex and a $q$
from the adjacent vertex obtains ${\bf p}_\perp = {\bf k}_{\perp j+1}-{\bf
k}_{\perp j}$. 

In case there are two or more identical bosons the matrix element
should be symmetrised and in general we obtain the
symmetrised production amplitude
\begin{eqnarray}
{\cal M} = \sum_{{\cal P}} {\cal M}_{{\cal P}}
\end{eqnarray}
where the sum goes over all possible permutations of the identical particles.
The squared amplitude occurring in Eq (\ref{e:lundprob}) will then be
\begin{eqnarray}
\label{interference}
|{\cal M}|^2= \sum_{{\cal P}} |{\cal M}_{{\cal P}}|^2 
\left(1 + \sum_{{\cal P}^{\prime}\neq {\cal P}} 
\frac{2 \mbox{Re}({\cal M}_{{\cal P}}{\cal M}_{{\cal P}^{\prime}}^{\ast})}
{|{\cal M}_{{\cal P}}|^2 + |{\cal M}_{{\cal P}^{\prime}}|^2}\right)
\label{e:symM2}
\end{eqnarray}
JETSET will provide the outer sum in Eq (\ref{e:symM2}) by the
generation of many events but it is evident that the model predicts a
quantum mechanical interference weight, $w_{{\cal P}}$, for each given
final state characterised by the permutation ${\cal P}$:
\begin{eqnarray}
\label{weight}
w_{{\cal P}}= 1 + \sum _{{\cal P}^{\prime}\neq {\cal P}} 
\frac{2 \mbox{Re}({\cal M}_{{\cal P}}{\cal M}_{{\cal P}^{\prime}}^{\ast})}
{|{\cal M}_{{\cal P}}|^2+{\cal M}_{{\cal P}^{\prime}}|^2}
\end{eqnarray}
In the Lund Model we note in particular for the case exhibited in Fig
\ref{f:lundarea}, with two identical bosons denoted 1 and 2 having a 
state $I$ in between, that the decay area is different if the two
identical particles are exchanged. It is evident that the
interference between the two permutation matrices will contain the
area difference, $\Delta A$, and the resulting general weight formula will
be
\eqbe 
w_{{\cal P}} = 1+\sum_{{\cal P}^{\prime}\neq {\cal P}}\frac {\textstyle
 \cos \frac{\textstyle \Delta A}{\textstyle 2\kappa}} {\textstyle
 \cosh \left( \frac{\textstyle b\Delta A}{\textstyle 2}+
 \frac{\textstyle \Delta(\sum {\bf k}^{2}_{\perp j})}{\textstyle
 2\kappa}
\right)}
\label{e:weight} 
\eqen
where $\Delta$ stands for the difference between the configurations
described by the permutations ${\cal P}$ and ${\cal P}^{\prime}$ and
the sum is taken over all the vertices. In our MC implementation of
the weight we replace the string constant $\kappa$ in the transverse
momentum generation with the default (in JETSET) transverse width,
$2\sigma^2$ (which is of the order of $\kappa$). The calculation of
the weight function for $n$ identical bosons contains $n!-1$ terms and
it is therefore from a computational point of view of
exponential-type. We have in \cite{r:bamr} introduced approximate
methods reducing it to power-type instead and we refer for
details to this work.

We have seen that the transverse and longitudinal components of the
particles momenta stem from different generation mechanisms. This is
clearly manifested in the weight in Eq~(\ref{e:weight}) where they
give different contributions. In the following we will therefore in
some detail analyse the impact of this difference on the transverse
and longitudinal correlation lengths, as implemented in the model.

In order to understand the properties of the weight in Eq
(\ref{e:weight}) we again consider the simple case in Fig
\ref{f:lundarea}.  The area difference of the two configurations
depends upon the energy momentum vectors $p_1, p_2$ and $p_I$ and can
in a dimensionless and useful way be written as
\eqbe
\frac{\Delta A}{2\kappa} = \delta p \delta x_L 
\label{e:deltaA}
\eqen
where $\delta p = p_2-p_1$ and $\delta x_{L} = (\delta t; 0, 0,
\delta z)$ is a reasonable estimate of the space-time difference,
along the surface area, between the production points of the two
identical bosons.

In order to preserve the transverse momenta of the particles in the
state $(1,I,2)$ it is necessary to change the generated
${\bf k}_{\perp}$ at the two internal vertices around the state $I$
during the permutation, i.e. to change the Gaussian weights.  Also in
this case we may write a formula similar to Eq (\ref{e:deltaA}) for
the transverse momentum change:
\begin{eqnarray}
\frac{\Delta (\sum {\bf k}_{\perp j}^2)}{2 \kappa}= \delta {\bf p}_\perp
\delta {\bf x}_\perp
\label{e:deltakt}
\end{eqnarray}
where ${\delta {\bf p}}_\perp$ is the difference ${\bf p}_{\perp
2}-{\bf p}_{\perp 1}$ and $\delta {\bf x}_\perp = ({\bf k}_{\perp b} -
(-{\bf k}_{\perp a}))/\kappa$. The two neighbouring vertices of the state
$(1,I,2)$ ($(2,I,1)$) are denoted by $a$ and $b$ and ${\bf k}_{\perp b} +{\bf
k}_{\perp a}$ corresponds to the states transverse momentum exchange
to the outside. Therefore $\delta {\bf x}_\perp$ constitutes a
possible estimate of the transverse distance between the production
points of the pair.

For the general case when the permutation ${\cal P}^{\prime}$ is more
than a two-particle exchange there are formulas similar to Eqs
(\ref{e:deltaA}) and (\ref{e:deltakt}) although they are more
complex (and the expressions do not vanish when only two of the
exchanged particles have the same energy momentum).

It is evident from the considerations leading to Eqs (\ref{e:deltaA})
and (\ref{e:deltakt}) that only particles with a finite longitudinal
distance and small relative energy momenta will give significant
contributions to the weights. We also note that we are in this way
describing longitudinal correlation lengths along the colour fields,
inside which a given flavour combination is compensated. The
corresponding transverse correlation length describes the tunneling
(and in this model it provides a damping chaoticity).

The weight distribution we obtain is discussed in \cite{r:bamr} (and
with varying kaon and baryon content also below). It is strongly
centered around unity although there are noticeable tails to
both larger and smaller (even negative) weights. The total production
probability is, however, positive and we find negligible changes in
the JETSET default observables (besides the correlation functions) by
this extension of the Lund model.

\section{Results} \vspace{-2ex}
Two-dimensional Bose-Einstein correlations in $e^+e^-$ annihilation have
been analysed at lower energies than LEP by the TASSO collaboration
\cite{r:tasso}. Although they find that their data is compatible with
a spherically symmetric correlation function they conclude that at
least one order of magnitude of more data is required to obtain more
detailed information. With the large statistics available from LEP we
have therefore generated $q\bar{q}$-events at the $Z^0$ pole to
investigate the properties of our model. Short-lived resonances like
the $\rho$ and $K^{*}$ are allowed to decay before the
BE-symmetrisation, while more long-lived ones are not affected.

We have analysed two-particle correlations in the Longitudinal
Centre-of-Mass System ({\it LCMS}). For each pair of particles the {\it LCMS} is
the system in which the sum of the two particles momentum components
along the jet axis is zero, which of course also means that the sum of
their momenta is perpendicular to the jet axis. The transverse and
longitudinal momentum differences are then defined in the {\it LCMS} as
\eabe
q_t = \sqrt{(p_{x2}-p_{x1})^2+(p_{y2}-p_{y1})^2} \\ \nonumber
q_L = |p_{z2}-p_{z2}| 
\eaen
where the jet axis is along the z-axis.

\begin{figure}[t]
  \hbox{\vbox{
\begin{center}
    \mbox{\psfig{figure=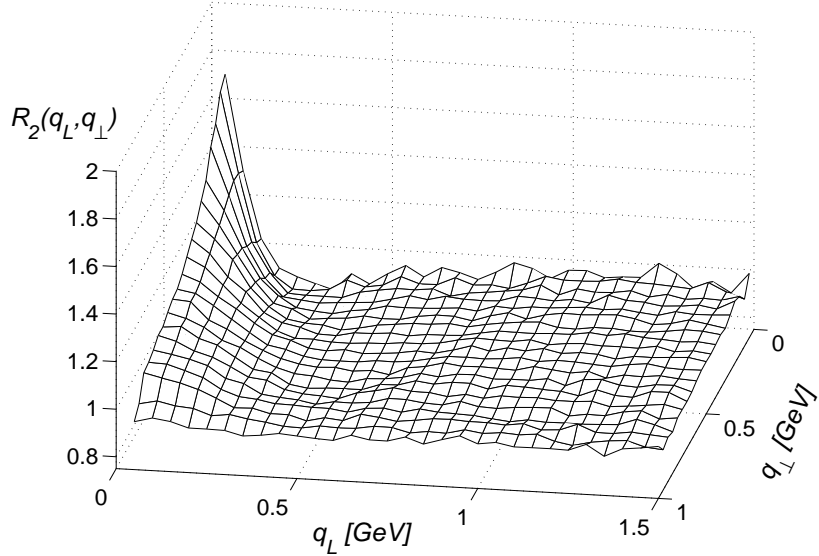,width=11cm}}
    \end{center} }} 
\caption{\em The ratio $R_{2}(q_L,q_\perp)$ of the number of charged 
pion pairs having relative four-momentum components $q_L$ and $q_\perp$ with and without Bose-Einstein weights applied. The sample consists of particles which are either initially produced or stemming from short-lived resonances.} 
\label{f:R2_2d} 
\end{figure}

We have taken the ratio of the two-particle probability density of pions, $\rho_2$, with and without BE weights applied as the two-particle correlation function, $R_2$
\eqbe
R_2(p_1,p_2)=\frac{\rho_{2w}(p_1,p_2)}{\rho_2(p_1,p_2)}
\label{e:R2}
\eqen
and the resulting function is shown in Fig~\ref{f:R2_2d}. It is
clearly seen that it is not symmetric in $q_L$ and $q_\perp$ and in
particular that the correlation length, as measured by the inverse of
the width of the correlation function, is longer in the longitudinal
than in the transverse direction. This difference remains for
reasonable changes of the width in the transverse momentum generation.
For comparison we have also analysed events where the Bose-Einstein
effect has been simulated by the LUBOEI algorithm implemented in
JETSET \cite{r:llts}. In LUBOEI the BE effect is simulated as a
mean-field potential between identical bosons which is spherically
symmetric in $Q$. Analysing only the initial particles and particles
stemming from short-lived decays results for the LUBOEI events in a
correlation function with identical transverse and longitudinal
correlation lengths. The correlation lengths are in agreement with the
source radii input to LUBOEI. Using all the final pion pairs, after
all decays, in the analysis results in a small decrease in the
transverse correlation length and of course a large decrease in the
height for $q_L\simeq q_\perp \simeq 0$, while the longitudinal
correlation length is rather unaffected. The pions from long lived
decays affect the correlation lengths in the same way both for our
model and for LUBOEI.

In \cite{r:bamr} it is shown that our model gives rise to genuine
three-particle correlations. We will in this letter continue to
investigate three-particle correlations and we will in particular use
our knowledge of the different contributions to the weight function to
study the genuine higher order correlations. We will also exhibit how
the genuine higher order terms in the weight function mainly clusters
particles in the longitudinal direction. 

The total three-particle correlation function is in analogy with Eq~(\ref{e:R2})
\eqbe
R_3^{''}(p_1,p_2,p_3)=\frac{\rho_{3w}(p_1,p_2,p_3)}{\rho_3(p_1,p_2,p_3)}
\eqen
To get the genuine three-particle correlation function, $R_3$, the
consequences of having two-particle correlations in the model have to
be subtracted from $R_3^{''}$. To this aim we have calculated the
weight taking into account only configurations where pairs are
exchanged, $w^{'}$. In this way the three-particle correlations which
only are a consequence of lower order correlations can be defined as
\eqbe
R_3^{'}(p_1,p_2,p_3)=\frac{\rho_{3w^{'}}(p_1,p_2,p_3)}{\rho_3(p_1,p_2,p_3)}
\eqen
The genuine three-particle correlation function, $R_3$, is then given by
\eqbe
R_3=R_3^{''}-R_3^{'}+1
\eqen
\begin{figure}[t]
  \hbox{\vbox{
\begin{center}
    \mbox{\psfig{figure=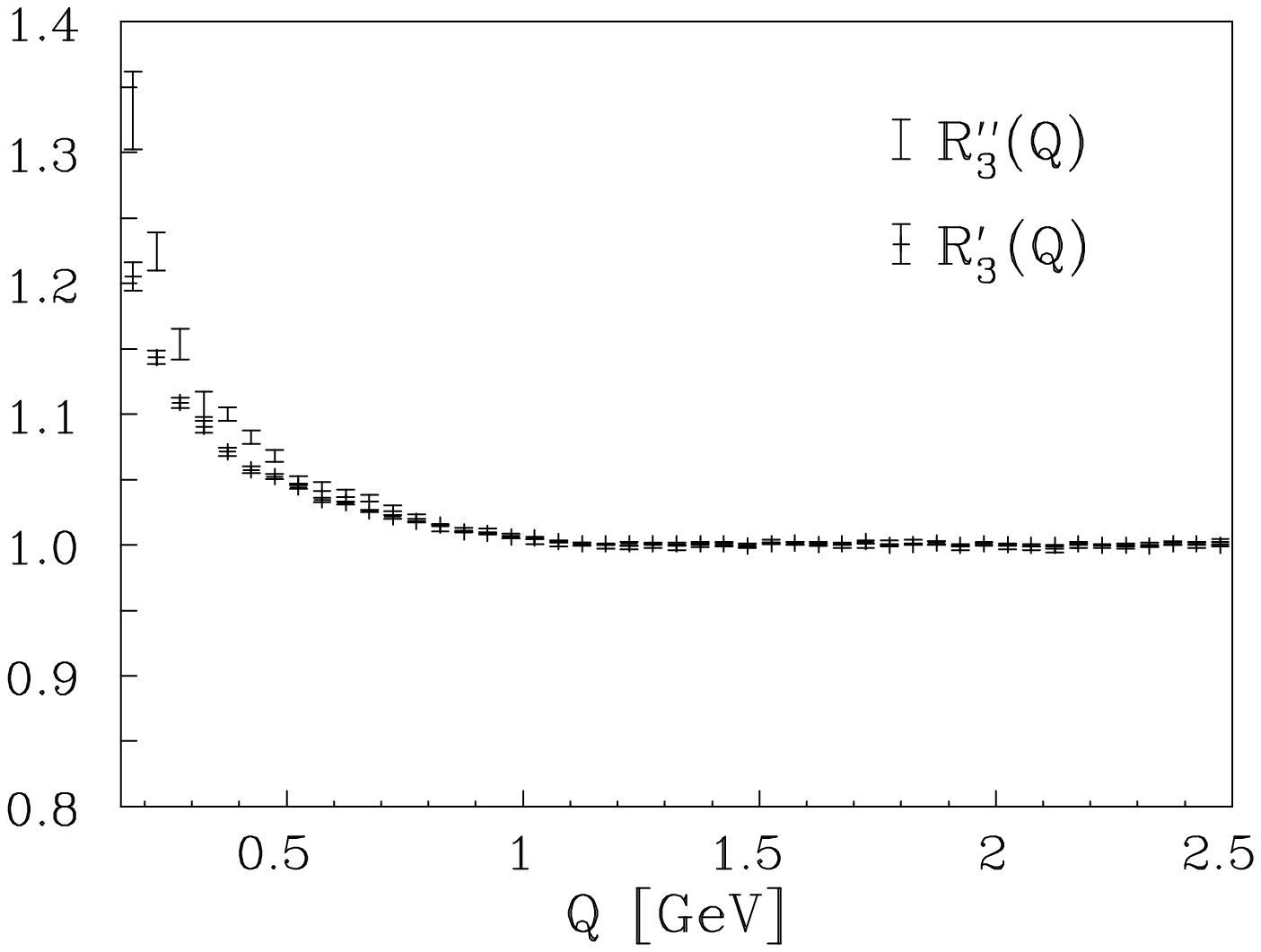,width=8cm}}
\mbox{\psfig{figure=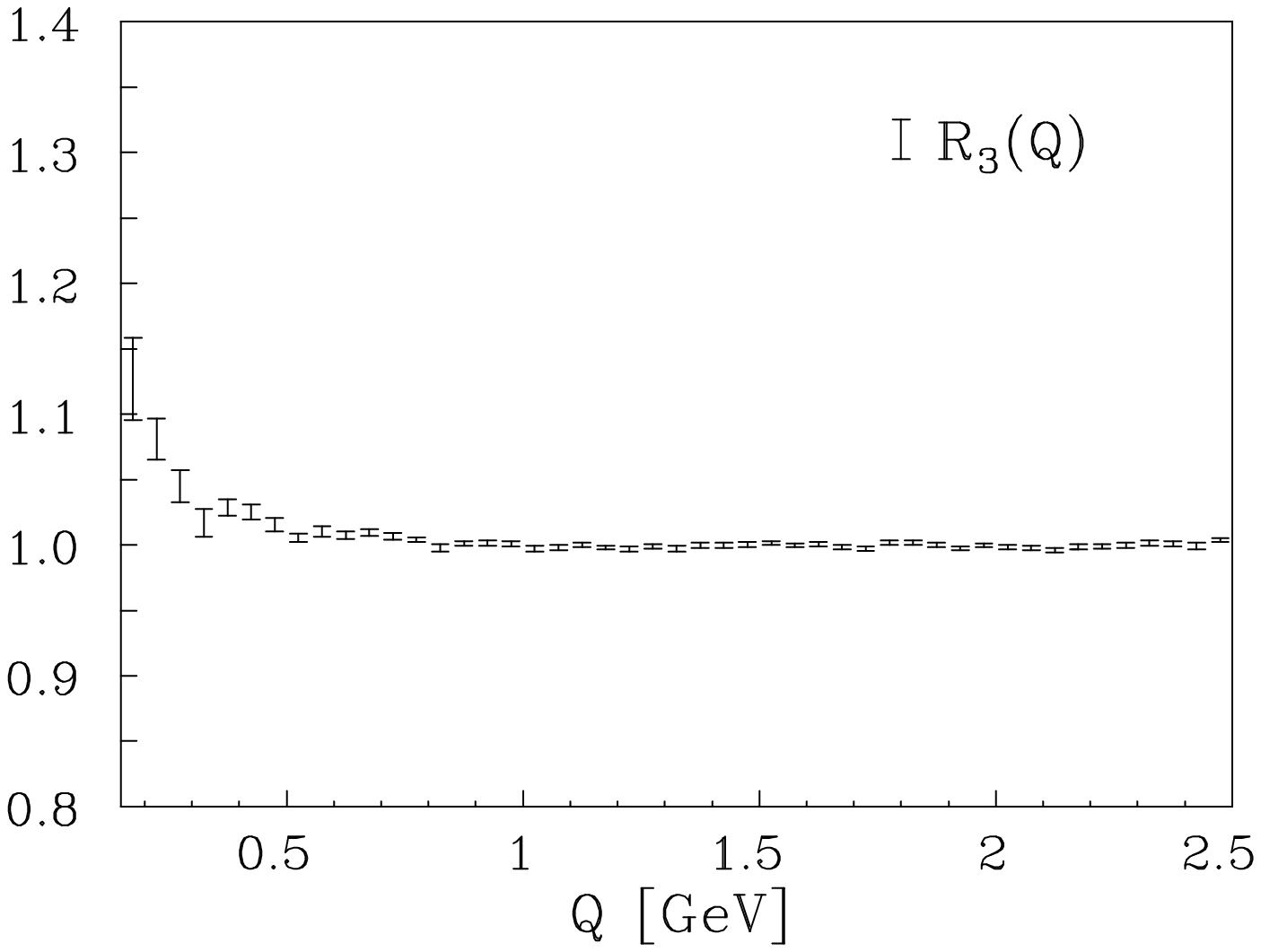,width=8cm}}
    \end{center} }} 
\caption{\em $R_3^{''}(Q)$ and $R_3^{'}(Q)$ are shown in the left figure, while the figure to the right shows $R_3(Q)$. The existence of genuine three-particle correlations is apparent.}
\label{f:R3} 
\end{figure}
We have analysed $R_3$ in one dimension as a function of the
kinematical variable
\eqbe
Q=\sqrt{Q_{12}^2+Q_{13}^2+Q_{23}^2} ~~~~\mbox{with}~~~~Q_{ij}^2=-(p_i-p_j)^2
\eqen
and in two dimensions we have used the following variables calculated
in the {\it LCMS} for each triplet of identical bosons
\eabe
q_L=\sqrt{q_{L12}^2+q_{L13}^2+q_{L23}^2} ~~~~\mbox{with}~~~~q_{Lij}^2=(p_{zi}-p_{zj})^2 \\
q_\perp=\sqrt{q_{\perp 12}^2+q_{\perp 13}^2+q_{\perp 23}^2} ~~~~\mbox{with}~~~~q_{\perp ij}^2=({\bf p}_{\perp i}-{\bf p}_{\perp j})^2 \nonumber
\eaen
where the $z$-axis is along the jet axis. 
\begin{figure}[t]
  \hbox{\vbox{
\begin{center}
    \mbox{\psfig{figure=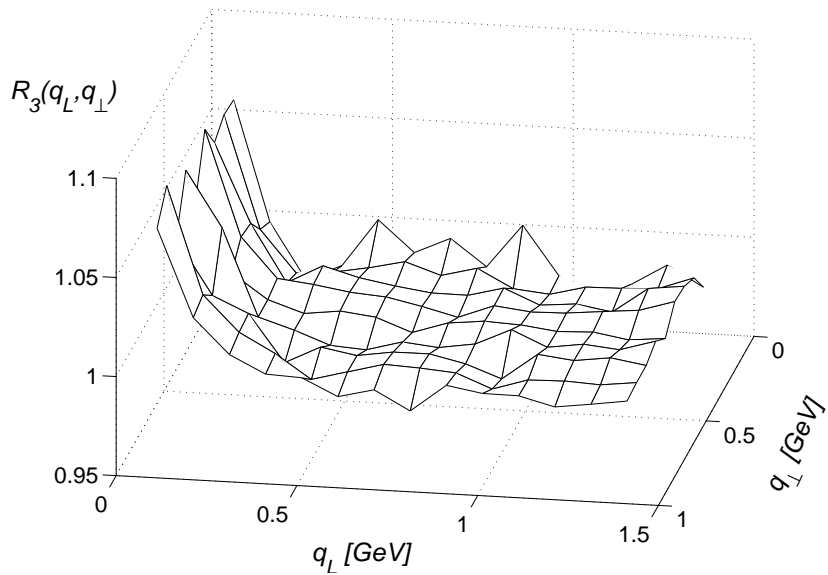,width=11cm}}
    \end{center} }} 
\caption{\em The ratio $R_{3}(q_L,q_\perp)$ of the number of triplets of 
charged pions with and without Bose-Einstein weights applied.} 
\label{f:R3_2d} 
\end{figure}
In Fig~\ref{f:R3} the correlation functions $R_3^{''}(Q), R_3^{'}(Q)$
and $R_3(Q)$ are shown, and the existence of genuine three-particle
correlations in the model is clearly exhibited. 

This way of getting the genuine correlations is not possible in an
experimental situation, where one has to find other ways to get a $R_3^{'}$
reference sample. We have suggested one possible option in
\cite{r:bamr} and the results in this letter are in agreement
with the conclusions of that investigation. In the present analysis
the contribution to the correlations from higher order configurations
in the weight calculation is apparent. We note that $R_3$ flattens out
earlier, i.e. for lower $Q$-values than $R_3^{''}$. This means that
the genuine three-particle correlations have a longer correlation
length compared to the consequences of lower order
correlations. Performing the same analysis in two dimensions in the
{\it LCMS} for each triplet results in the $R_3(q_L,q_\perp)$
distribution shown in Fig~\ref{f:R3_2d}. The effect of the higher
order terms is to pull the triplets closer in the longitudinal
direction while the transverse direction is rather unaffected.  This
suggests that higher order correlations are more sensitive to the
longitudinal stretching of the string field.

We have also studied the correlation length for pion pairs as a
function of the final charged multiplicity and the kaon content of the
state.  Within statistical errors which are relatively large we see no
dependence on either the charged multiplicity or the number of
kaons. Since one might suspect that events with many pions are
premiered by the re-weighting the average baryon and kaon content of
the events have been investigated. We find that the changes of the
average multiplicity of different kaon species as well as of the
average multiplicity of protons and neutrons in the final state are
much smaller than the experimental errors as summarised in~\cite{r:pdg}.

\end{document}